\documentclass[aps,prl,reprint,twocolumn,superscriptaddress,showpacs,notitlepage]{revtex4-1}

\usepackage{graphicx}
\usepackage{mathrsfs}
\usepackage{bm}
\usepackage{amsmath}
\usepackage{dcolumn}
\usepackage{epstopdf}
\usepackage{dsfont}
\usepackage{amssymb}
\usepackage{tabularx}
\usepackage{array}
\usepackage{float}
\usepackage{color}
\usepackage{epstopdf}
\usepackage{mathrsfs}
\usepackage[colorlinks, linkcolor=blue,anchorcolor=blue,citecolor=blue,urlcolor=blue]{hyperref}

\begin{document}
\title{Symmetry-Enforced Nodal Chain Phonons}

\author{Jiaojiao Zhu}
\address{Research Laboratory for Quantum Materials, Singapore University of Technology and Design, Singapore 487372, Singapore}

\author{Weikang Wu}\email{weikang.wu@ntu.edu.sg}
\address{Division of Physics and Applied Physics, School of Physical and Mathematical Sciences, Nanyang Technological University, Singapore, 637371, Singapore}
\address{Research Laboratory for Quantum Materials, Singapore University of Technology and Design, Singapore 487372, Singapore}

\author{Jianzhou Zhao}\email{jzzhao@swust.edu.cn}
\address{Research Laboratory for Quantum Materials, Singapore University of Technology and Design, Singapore 487372, Singapore}
\address{Co-Innovation Center for New Energetic Materials, Southwest University of Science and Technology, Mianyang 621010, China}

\author{Hao Chen}
\address{NNU-SULI Thermal Energy Research Center (NSTER) \& Center for Quantum Transport and Thermal Energy Science (CQTES),
School of Physics and Technology, Nanjing Normal University, Nanjing 210023, China}

\author{Lifa Zhang}
\address{NNU-SULI Thermal Energy Research Center (NSTER) \& Center for Quantum Transport and Thermal Energy Science (CQTES),
School of Physics and Technology, Nanjing Normal University, Nanjing 210023, China}

\author{Shengyuan A. Yang}
\address{Research Laboratory for Quantum Materials, Singapore University of Technology and Design, Singapore 487372, Singapore}


\begin{abstract}
Topological phonons in crystalline materials have been attracting great interest. However, most cases studied so far are direct generalizations of the topological states from electronic systems. Here, we reveal a novel class of topological phonons---the symmetry-enforced nodal-chain phonons,  which manifest features unique for phononic systems. We show that with $D_{2d}$ little co-group at a non-time-reversal-invariant-momentum point, the phononic nodal chain is guaranteed to exist owing to the vector basis symmetry of phonons, which is a unique character distinct from electronic and other systems. Combined with the spinless character, this makes the proposed nodal-chain phonons enforced by symmorphic crystal symmetries. We further screen all 230 space groups, and find five candidate groups. Interestingly, the nodal chains in these five groups exhibit two different patterns: for tetragonal systems, they are one-dimensional along the fourfold axis; for cubic systems, they form a three-dimensional network structure. Based on first-principles calculations, we identify K$_{2}$O as a realistic material hosting almost ideal nodal-chain phonons. We show that the effect of LO-TO splitting, another unique feature for phonons, helps to expose the nodal-chain phonons in K$_{2}$O in a large energy window. In addition, all the five candidate groups have spacetime inversion symmetry, so the nodal chains also feature a quantized $\pi$ Berry phase. This leads to drumhead surface phonon modes that must exist on multiple surfaces of a sample.

\end{abstract}

\maketitle

Topological quasiparticles, emerged around band degeneracy points in condensed matters, have been attracting tremendous research interest in the past decade. The field was initiated with the focus on electronic systems. Pioneering examples include electrons around Weyl~\cite{weyl-electron-1,weyl-electron-2,weyl-electron-3,weyl-dirac-electron} and Dirac~\cite{dirac-electron-1,dirac-electron-2,dirac-electron-3,dirac-electron-4,weyl-dirac-electron} points in the band structures, which resemble Weyl and Dirac fermions and thereby can simulate fascinating effects from high energy physics~\cite{high-energy-physics-1,high-energy-physics-2,high-energy-physics-3}. Moreover, condensed matter systems respect the space group (SG) symmetry, which is a much smaller subgroup of the Poincar$\acute{\mbox{e}}$ symmetry. The reduced constraints permit a rich variety of novel emergent quasiparticles beyond the Weyl/Dirac paradigm ~\cite{beyond-1,beyond-2,beyond-3}. For instance, band degeneracies may form higher-dimensional manifolds in the momentum space, leading to
nodal-line~\cite{nodalline-electron-1,dirac-electron-3, nodalline-electron-2,nodalline-electron-3,nodalline-electron-4,nodalline-electron-5,nodalline-electron-6,nodalline-electron-7,nodalline-electron-8,nodalline-electron-9,nodalline-electron-10} and even nodal-surface~\cite{nodalsurface-electron-1,nodalsurface-electron-2,nodalsurface-electron-3,nodalsurface-electron-4} electrons, with unique topological boundary modes and effects.

It was later realized that this research can be naturally extended to bosonic and even classical systems. Particularly, there is a surge of interest recently in exploring novel quasiparticles in phonons~\cite{phonons-1,phonons-2,phonons-3,phonons-4,phonons-5,phonons-6}, which describe the atomic lattice vibrations in solids. This is also motivated by the advance in experimental techniques which can now probe the full THz phonon spectrum with meV-resolution~\cite{mev-1,mev-2,mev-3}. A number of materials with Weyl, Dirac, and nodal-line phonons have been predicted~\cite{weyl-phonon-1,triple-point,weyl-phonon-2,nodalline-phonon-1,weyl-phonon-3,weyl-phonon-4,weyl-phonon-5,dirac-phonon-2}, and some have been successfully verified in experiment~\cite{nodalline-phonon-1,ixs}. Nevertheless, as direct extensions of concepts from electronic systems, except for the particles statistics, these phonons share essentially the same features as their electronic counterparts.

The question is: \emph{Can we find novel topological phonons with features unique for phononic systems?} In this work, we present such an example --- the symmetry-enforced nodal-chain phonons.

\begin{figure}[tb!]
    \centering
    \includegraphics[width=1.0\linewidth]{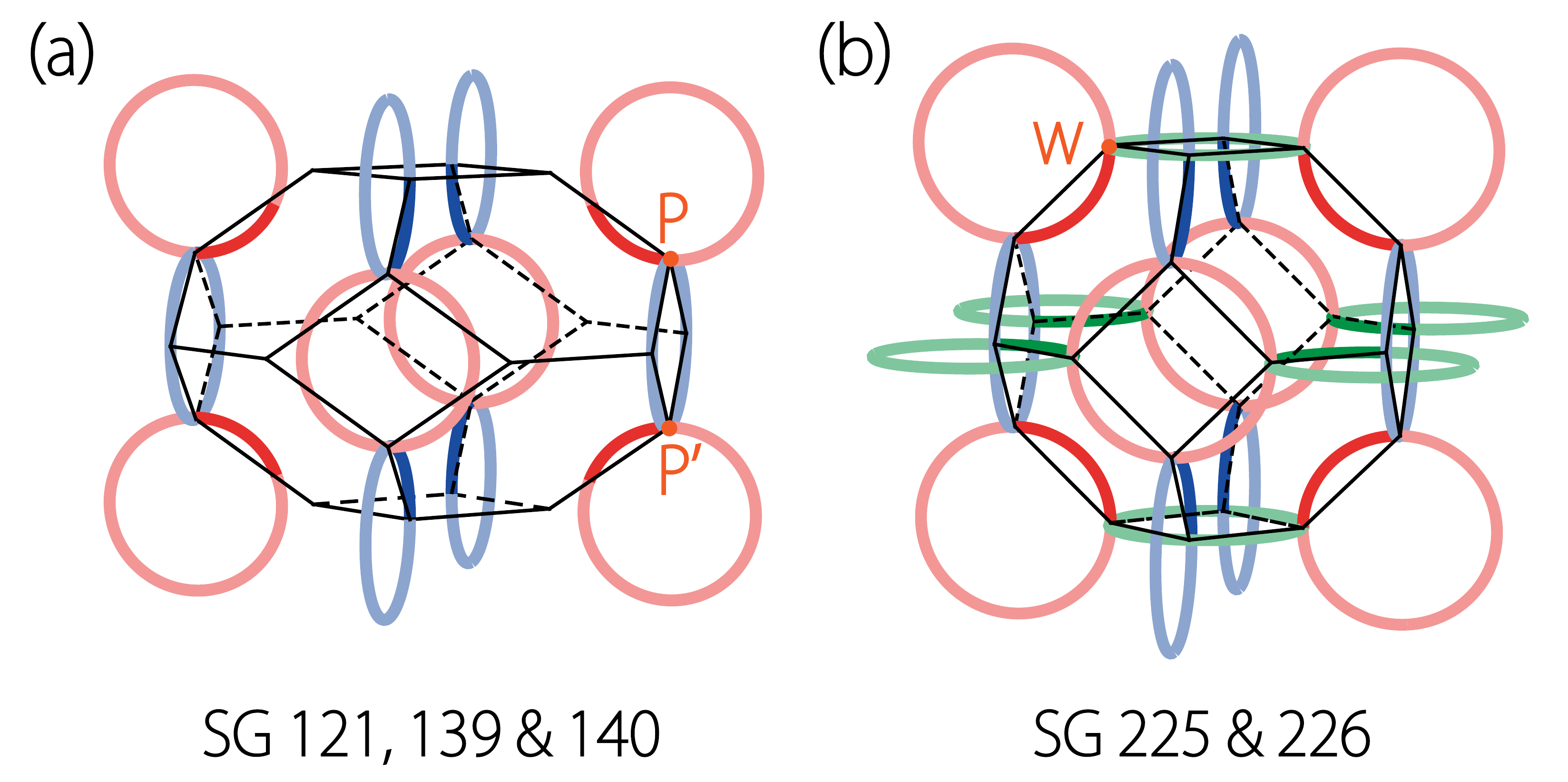}
    \caption{Symmetry-enforced nodal chains for the five candidate SGs. (a) SG 121, 139 and 140 host nodal chains running along the fourfold $S_{4z}$ axis.  (b)  In SG 225 and 226, there are chains running along all three directions, forming a network structure. Here, the rings with different orientations are marked with different colors. }
    \label{fig:1}
\end{figure}

A nodal chain is composed of multiple nodal rings touching at isolated points and is extended in momentum space (e.g., see Fig.~\ref{fig:1}). The concept was initially studied in electronic systems, where the electronic nodal chains usually require complicated nonsymmorphic crystal symmetries to be robust against spin-orbit coupling (SOC) ~\cite{non-centrosymmetric-nodalchain-electron-1,non-centrosymmetric-nodalchain-electron-2}. Nodal chains with symmorphic symmetries were also discussed~\cite{centrosymmetric-nodalchain-electron-1,centrosymmetric-nodalchain-electron-2,centrosymmetric-nodalchain-electron-3}, but they are typically destroyed by SOC, and more importantly, they are \emph{not} symmetry enforced, meaning that their presence in the spectrum depends on the system details and is not guaranteed. Clearly, these features pose obstacles for the search of electronic nodal chains in real materials.

In contrast, we find a class of nodal-chain phonons that are \emph{enforced} by symmorphic symmetries. We show that such phonons are dictated by the presence of $D_{2d}$ little co-group at non-time-reversal-invariant-momentum (non-TRIM) points. The key point is that unlike electrons and other systems, where the basis at lattice sites can take different symmetries (e.g., $s$, $p$, $d$, etc.), for phonons, the lattice displacement at each site is a vector, hence the basis is constrained to have the $p$-orbital symmetry. Under this vector representation, the nodal chain is guaranteed by the above-mentioned symmorphic symmetry. In addition, distinct from electrons, the vulnerability under SOC is not an issue here, since phonons are intrinsically spinless.
In this sense, the proposed symmetry-enforced nodal chain is indeed unique for phononic systems. We screen through the 230 SGs and find 5 groups that host such nodal chain phonons. Interestingly, depending on the SG, there are two different nodal-chain patterns, as shown in Fig.~\ref{fig:1}. Guided by the symmetry condition, we propose that K$_2$O, an existing material, is a candidate with almost ideal nodal chain phonons. Particularly, we show that the LO-TO splitting~\cite{lo-to}, which is another feature unique for phonons, helps to expose the phononic nodal chain in K$_2$O as the only band degeneracy in a large frequency window, facilitating the experimental detection. In addition, due to the spacetime inversion symmetry $\mathcal{PT}$, each ring in the chain enjoys an additional protection by the $\pi$ Berry phase~\cite{zak}, which also leads to the protected drumhead surface phonon modes~\cite{dirac-electron-3,nodalline-electron-2}.

\begin{figure}[tb!]
    \centering
    \includegraphics[width=1.0\linewidth]{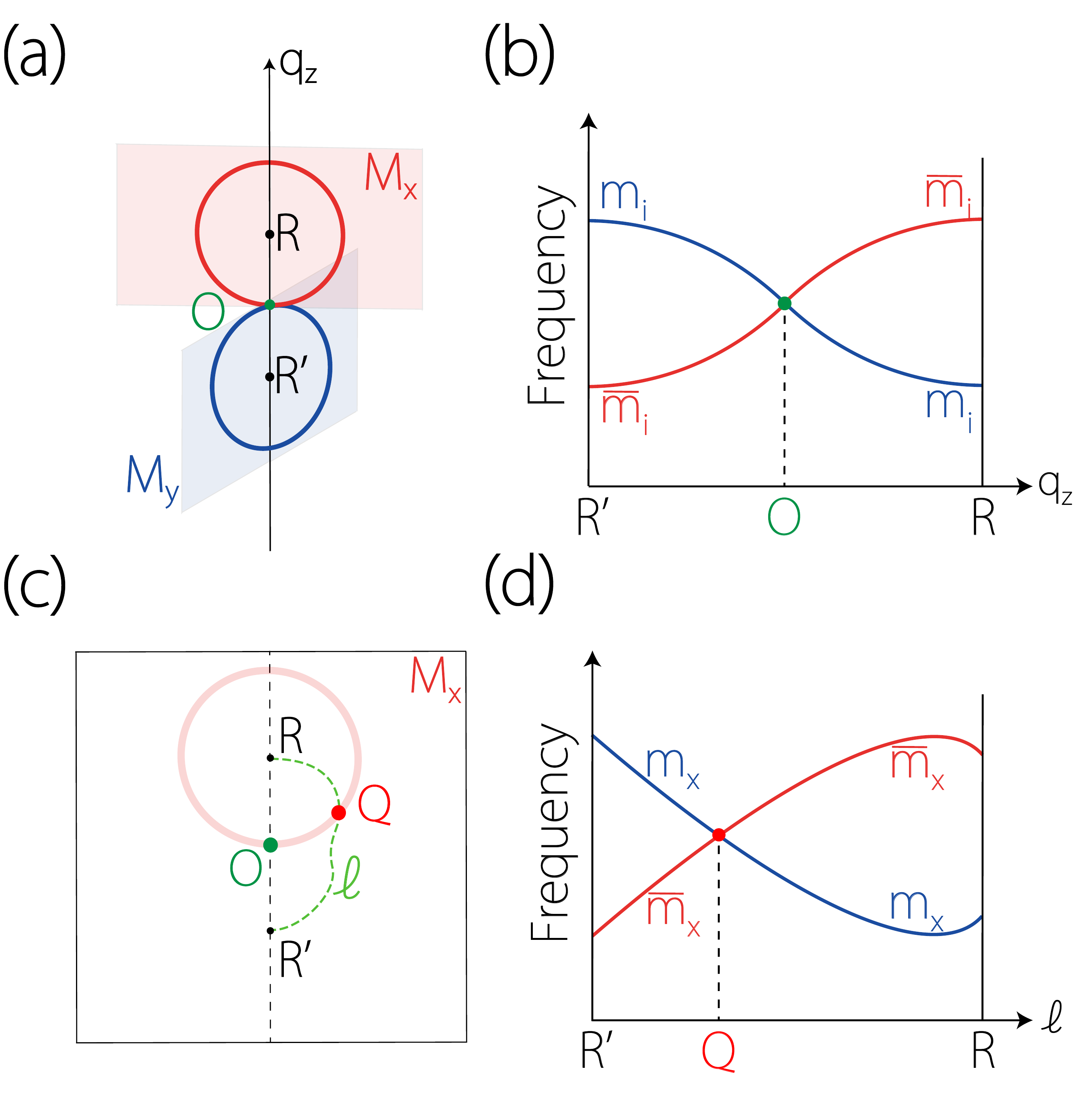}
    \caption{(a) Schematic diagram for a nodal chain formed by two rings located in two mutually orthogonal mirror planes $M_{x}$ and $M_{y}$. $O$ is a non-TRIM point with $D_{2d}$ little co-group.  (b) Spectrum along the $q_z$ axis for a pair of phonon branches that correspond to the $E$ representation of $D_{2d}$ at $O$. $m_i$ ($i=x,y$) are the mirror eigenvalues. (c) An arbitrary path $\ell$ in the $M_x$ plane, which connects a point $R$ on the $q_z$ axis to its $S_{4z}$ image $R'$.  The spectrum along $\ell$ is shown in (d), where the two branches must still cross at some point $Q$, as they have opposite $m_x$ eigenvalues. The crossing point traces out a ring in the $M_x$ plane, as shown in (c).}
    \label{fig:2}
\end{figure}

{\color{blue}\textit{Symmetry condition.}} We propose that the $D_{2d}$ group at a non-TRIM point of the Brillouin zone (BZ) enforces nodal-chain phonons. The $D_{2d}$ group contains two mutually orthogonal mirrors, which we may denote as $M_x$ and $M_y$, and a fourfold roto-reflection $S_{4z}$, which connects the two mirrors. Assume that $D_{2d}$ is the little co-group at a non-TRIM point $O$ in the BZ, and the $q_z$ axis through $O$ corresponds to the $S_{4z}$ axis, as shown in Fig.~\ref{fig:2}(a).

The eigenstates at $O$ correspond to the irreducible representations of the $D_{2d}$ group. Importantly, for phonons, as we have mentioned, the basis symmetry is constrained to be vectors, such that the vibrations normal to the principal ($S_{4z}$) axis, which have $p_x$ and $p_y$ basis symmetry, must constitute the two-dimensional irreducible representation $E$ of $D_{2d}$. It follows that such phonon branches must form twofold degenerate pairs at $O$, corresponding to the $E$ representation.

Let us consider one such degenerate pair at $O$. Since $[M_x, M_y]=0$, the two states can be chosen as simultaneous eigenstates of the two mirrors, and they
must have opposite $M_x$ as well as $M_y$ eigenvalues. If we denote one state as $|m_x,m_y\rangle$ with $m_{x/y}\in\{+1,-1\}$ the mirror eigenvalues, then the other state must be $|\overline{m}_x,\overline{m}_y\rangle$, where $\overline{m}_i\equiv -m_i$. In the subspace of this pair, the symmetry operations satisfy the following relations
\begin{equation}\label{Erep}
   \{\mathcal{M}_x, \mathcal{S}_{4z}\}=0,\quad \{\mathcal{M}_y, \mathcal{S}_{4z}\}=0.
\end{equation}
Here, the script symbols denote the symmetry operators represented in the subspace.

When moving along the $q_z$ axis away from $O$, the degenerate pair will generally split.
However, the symmetries $M_x$ and $M_y$ are preserved on the $q_z$ axis. Consider the spectrum of the two branches on the $q_z$ axis. First, it must be symmetric with respect to $O$, since the spectrum at point $(0,0,q_z)$ is connected to that at $(0,0,-q_z)$ by $S_{4z}$. Second, due to Eq.~(\ref{Erep}), the two states connected by $S_{4z}$ at $(0,0,\pm q_z)$ with the same energy must have opposite $m_x$ and $m_y$ values. This is schematically illustrated in Fig.~\ref{fig:2}(b), showing that the two branches with opposite $m_i$ ($i=x,y$) cross each other at $O$.

Now, consider an arbitrary path $\ell$ in the $M_x$ plane, which connects a point $R$, say at $(0,0,\pi/2)$, to its $S_{4z}$ image $R'$ at $(0,0,-\pi/2)$ [see Fig.~\ref{fig:2}(c)]. Since $\ell$ is in the $M_x$ plane, $m_x$ for states on the path is still well defined. Meanwhile, because the $m_x$ eigenvalues are flipped between $R$ and $R'$, the two phonon branches that we studied above must also cross each other at some point $Q$ on the path $\ell$, as illustrated in Fig.~\ref{fig:2}(c,d). Since $\ell$ is arbitrary, the crossing point must trace out a nodal ring passing through $O$ and lying in the $M_x$ plane. The same argument applies to the $M_y$ plane and results in another nodal ring. The two rings are perpendicular to each other, touch at point $O$ [and also at $(0,0,\pm\pi)$ which is another $D_{2d}$-invariant point], and are connected by $S_{4z}$. Therefore, they compose a nodal chain in the momentum space running along the $S_{4z}$ axis, as illustrated in Fig.~\ref{fig:2}(a).

We have a few remarks. First, it is clear that the unique vector basis symmetry of phononic systems plays a crucial role here, which, combined with the symmorphic $D_{2d}$ symmetry, enforces the presence of nodal chain phonons. Although symmetry cannot determine the energy of the chain, owing to the bosonic character and the experimental technique such as inelastic x-ray scattering (IXS) that can probe the whole THz phonon spectrum ~\cite{nodalline-phonon-1,ixs}, the experimental detection should not be an issue. In comparison, for electronic systems, the existence of similar kind of chain cannot be guaranteed, especially in the limited window around Fermi energy that we can probe in experiment.

Second, the analysis above applies only for spinless particles. For spinful ones like electrons, the SOC will generally destroy the nodal chain ~\cite{centrosymmetric-nodalchain-electron-1}. Together with the first point, one can see that the proposed chain indeed manifests the characteristics of phononic systems.

Third, we required the point $O$ to be a non-TRIM point. The reason is that if it is a TRIM point, then the time reversal symmetry $\mathcal{T}$ and $S_{4z}$ would require the two branches that are degenerate at $O$ remain degenerate on the $q_z$ axis. In this case, we have only a single phononic nodal line, rather than a chain.

Fourth, the touching of two perpendicular rings at the $O$ point can also be inferred from the $k\cdot p$ effective model expanded at $O$. Constrained by the $D_{2d}$ group and in the $E$ basis, we obtain the following effective model expanded to $q$-quadratic order
\begin{equation}
  \mathcal{H}_\text{eff}(\bm q)= \epsilon(\bm{q})\sigma_{0} + c_{1}q_{x}q_{y}\sigma_{x} + [c_{2}q_{z} + c_{3}(q_{x}^{2} - q_{y}^{2})]\sigma_{z},
\end{equation}
where $\epsilon(\bm{q}) = \varepsilon_{1}(q_{x}^{2} + q_{y}^{2}) + \varepsilon_{2}q_{z}^{2}$, $\varepsilon_{i}$ and $c_{i}$ are real model parameters, and $\sigma$'s are Pauli matrices. The  model shows a linear band splitting along $q_{z}$ and quadratic splitting in the $q_{x}$-$q_{y}$ plane. The degeneracy manifold indeed conforms with that of two orthogonal nodal rings (see details in the Supplemental Material).

\begin{table}
\renewcommand\arraystretch{1.5}
\caption{\label{123} SGs hosting the proposed symmetry-enforced nodal-chain phonons. Here, the coordinates of the ring touching points ($O$ point in Fig.~2(a)) are also provided.}
\begin{tabular}{ccccc}
\hline\hline
SG No.  & \qquad Crystal system  & \qquad Touching point\\
\hline
121 & \qquad $\mathrm{Tetragonal}$  & \qquad $P$($ \frac{1}{4}$, $\frac{1}{4}$, $\frac{1}{4}$) $P^{\prime}$($-\frac{1}{4}$, $-\frac{1}{4}$, $\frac{3}{4}$)\\
139 & \qquad $\mathrm{Tetragonal}$  & \qquad $P$($\frac{1}{4}$, $\frac{1}{4}$, $\frac{1}{4}$) \\
140 & \qquad $\mathrm{Tetragonal}$  & \qquad $P$($\frac{1}{4}$, $\frac{1}{4}$, $\frac{1}{4}$)\\
225 & \qquad $\mathrm{Cubic}$  & \qquad $W$($\frac{1}{2}$, $\frac{1}{4}$, $\frac{3}{4}$)\\
226 & \qquad $\mathrm{Cubic}$  & \qquad $W$($\frac{1}{2}$, $\frac{1}{4}$, $\frac{3}{4}$) \\
\hline\hline
\end{tabular}
\end{table}

Following the symmetry condition, we examine all the 230 SGs and obtain 5 candidate groups that host the symmetry-enforced nodal-chain phonons.
These groups and the corresponding $O$ points are listed in Table~\ref{123}. The two different patterns of the nodal chains are illustrated in  Fig.~\ref{fig:1}. One observes that the chains in SG~121, 139, and 140 form a one-dimensional structure running along the $S_{4z}$ axis. In comparison, there are three families of orthogonal chains for SG~225 and 226, forming a three-dimensional chain network.

In addition, we note that all SGs in Table~\ref{123} possess the inversion symmetry $\mathcal{P}$. The combined $\mathcal{PT}$ symmetry enforces a $\pi$-quantized  Berry phase for arbitrary closed loops in momentum space, constituting an one-dimensional $\mathbb{Z}_2$ topological charge for $\mathcal{PT}$-invariant systems. This offers the nodal rings an additional protection, namely, each ring is protected by the $\pi$ Berry phase defined on a small loop encircling the ring. As a result, even when the symmetry is reduced by certain perturbations on the system, each ring should persist as long as $\mathcal{PT}$ is still preserved.

\begin{figure}[ht!]
    \centering
    \includegraphics[width=1.0\linewidth]{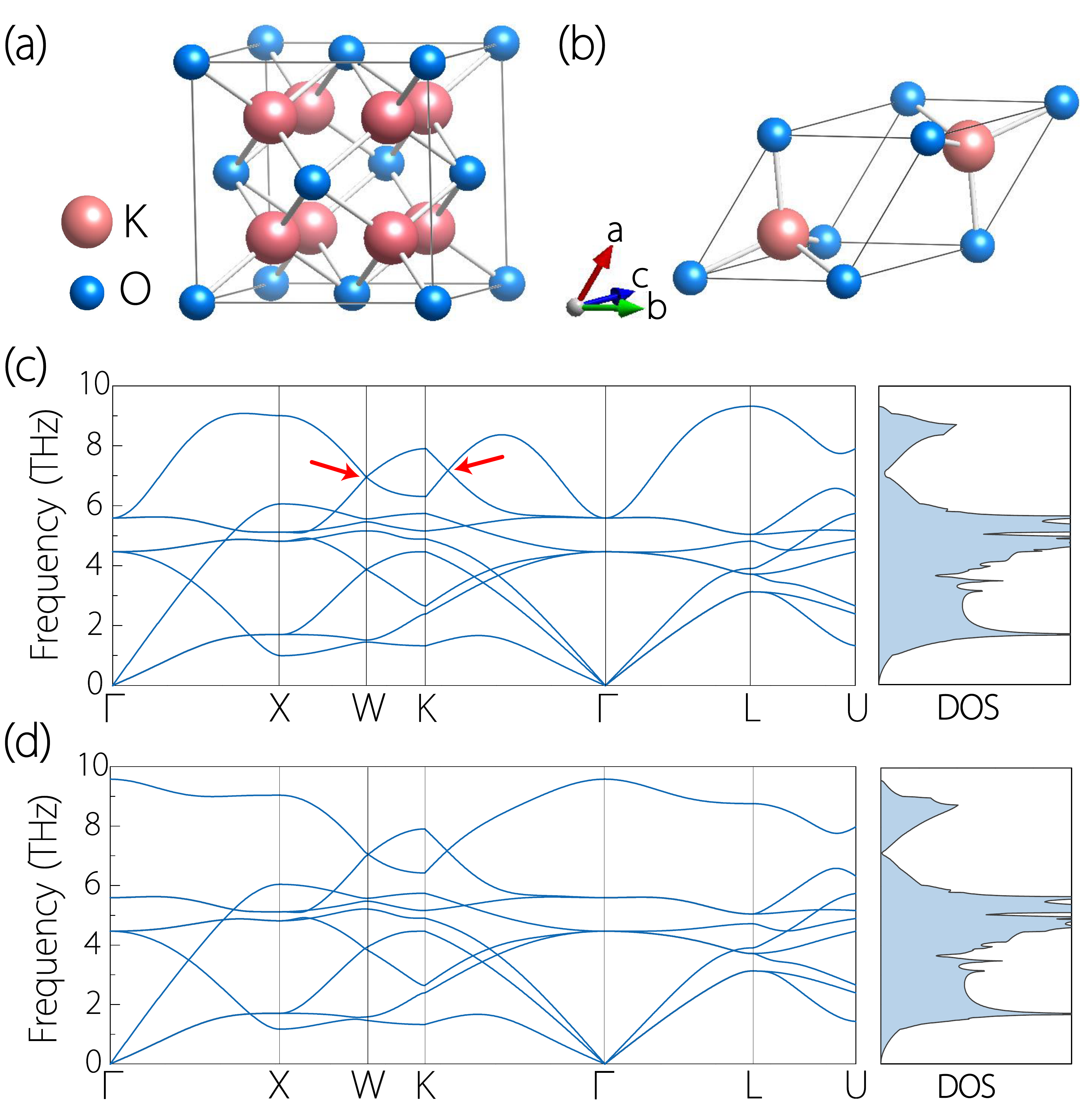}
    \caption{(a) Conventional unit cell and (b) primitive cell of $\mathrm{K_2O}$. (c) Calculated phonon spectrum and phonon density of states for  $\mathrm{K_2O}$. The arrows indicate points on the nodal chain. (d) Calculated phonon spectrum and phonon density of states with non-analytic correction (LO-TO splitting) included.}
    \label{fig:3}
\end{figure}

{\color{blue}\textit{Nodal-chain phonons in K$_2$O.}} Guided by the symmetry condition, we identify an existing material K$_{2}$O as a candidate with almost ideal nodal-chain phonons. The K$_{2}$O crystal was synthesized long ago in the 1930s~\cite{k2o}. As shown in Fig.~\ref{fig:3}(a), it has the antifluorite crystal structure with the SG~225 ($Fm\bar{3}m$), which is one of the candidates in Table \ref{123}. We investigate its properties by using the first-principles calculations based on the density functional theory (DFT). The calculation details are presented in the Supplemental Material. The optimized conventional lattice constant is $a = 6.49$ $\text{\AA}$ (Fig.~\ref{fig:3}(a)), which agrees well with the experimental value of $6.44$ $\text{\AA}$~\cite{k2o}. The two types of atoms K and O occupy the $8c$ and $4a$ Wyckoff positions, respectively.

The calculated phonon spectrum of K$_{2}$O is plotted in Fig.~\ref{fig:3}(c).  According to Table \ref{123} and Fig.~\ref{fig:1}(b), for SG~225, the ring touching point $O$ corresponds to the $W$ point of the BZ, and there are three mutually orthogonal rings, forming a network of chains in the extended BZ. In Fig.~\ref{fig:3}(c), a chain at around 7 THz formed by two optical branches can be clearly observed (indicated by the red arrows). A careful scan of the BZ confirms that the crossing between the top two branches form the chain pattern consistent with  Fig.~\ref{fig:1}(b).

Notably, in Fig.~\ref{fig:3}(c), the phonon band structure around the chain is not very ``clean", because the top phonon branch bends down near the BZ center. Fortunately, this is remedied by including the non-analytic correction from long-range Coulomb interactions, which is typically pronounced for ionic crystals such as K$_2$O. This correction leads to the well-know energy splitting between longitudinal optical and transverse optical phonon branches near the BZ center, i.e., the LO-TO splitting. In Fig.~\ref{fig:3}(d), one observes that the correction results in a large LO-TO splitting and pushes up the top branch. Meanwhile, the dispersion around the chain is more or less unaffected. Consequently, the phononic nodal chain is now well exposed in a large frequency window with a width $\sim 1$ THz. The energy of the chain can be readily inferred from the dip in the phonon density of states.

The clean band structure, the relatively large frequency window, and the small energy variation on the chain make K$_2$O an almost ideal candidate for experimental studies of nodal-chain phonons. The LO-TO splitting, which helps to further expose the nodal chain, is another unique feature for phononic systems, not present in electronic and other systems.

\begin{figure}[ht!]
    \centering
    \includegraphics[width=1.0\linewidth]{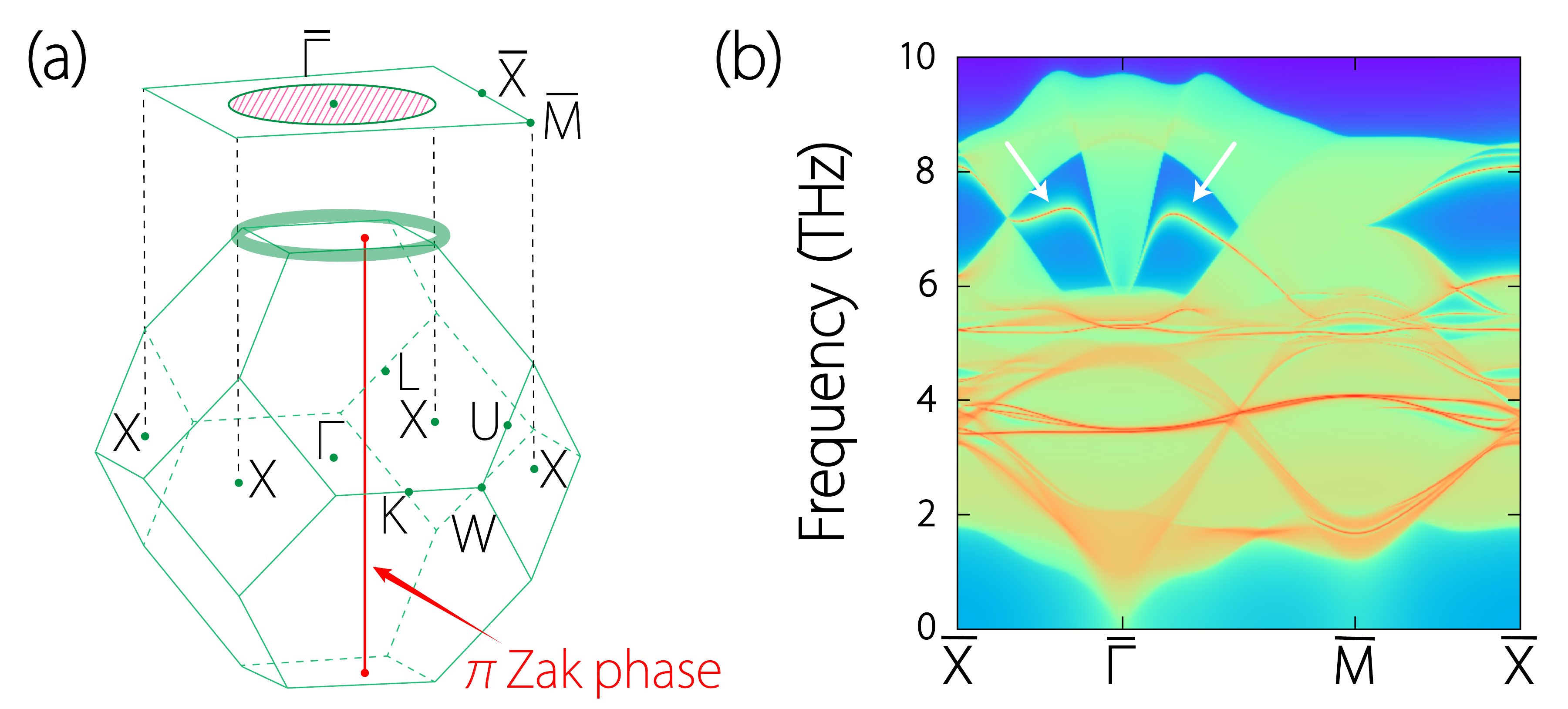}
    \caption{(a) K$_2$O's BZ and the corresponding (001) surface BZ. A straight path passing through the nodal ring has a $\pi$ Zak phase. This leads to protected drumhead surface modes. Here, we only show the ring in the horizontal plane. The analysis also applies to the other rings of the chain. (b) Calculated surface phonon spectrum for the (001) surface of K$_2$O (without non-analytic correction). The arrows indicate the drumhead surface phonon modes. }
    \label{fig:4}
\end{figure}

{\color{blue}\textit{Topological surface phonon modes.}} We have mentioned that owing to the $\mathcal{PT}$ symmetry, each ring of the chain features a quantized $\pi$ Berry phase. It follows that the Zak phase, defined as the Berry phase along a straight line traversing the BZ, must change by $\pi$ when the line crosses a ring. The $\pi$ Zak phase is verified by our first-principles calculations, as indicated in Fig.~\ref{fig:4}(a). It leads to the protected drumhead like surface modes, which span the region in the surface BZ bounded by the projection of the ring.

In Fig.~\ref{fig:4}(b), we plot the calculated surface phonon spectrum for the (001) surface of K$_2$O. One indeed finds the drumhead surface phonon modes, as indicated by the arrows. Drumhead surface modes also exist for conventional nodal-ring states, but they typically exist only on particular surfaces. For example, when the ring is parallel to the (001) surface, it is not going to produce drumhead surface modes on (100) and (010) surfaces. In comparison, since a nodal chain here is composed of orthogonal nodal rings, it must have drumhead surface modes simultaneously on multiple surfaces.

{\color{blue}\textit{Discussion.}} We have revealed a novel topological phonon state that manifests unique features of phononic systems, including the vector basis symmetry, the spinless nature, and the LO-TO splitting effect. These features help to enforce the existence of the phononic nodal chain and expose it in the spectrum, which are in sharp contrast to electronic systems.

We have provided detailed symmetry conditions for searching concrete material candidates. Because of the symmetry-enforced character, the search is expected to be extremely efficient. As an example, we have identified the material K$_2$O as a host of almost ideal nodal-chain phonons. In experiment, the bulk phonon dispersion can be imaged by inelastic x-ray scattering (IXS) ~\cite{mev-1,mev-2,mev-3} or neutron scattering~\cite{neutron}. The surface phonon modes can be probed by the high-resolution electron energy loss spectroscopy~\cite{high-resolution}, helium scattering~\cite{helium}, or THz spectroscopy~\cite{thz-1,thz-2}. Particularly, recent experiments with inelastic x-ray scattering have successfully mapped out topological phonons with meV-resolution, which is sufficient for detecting the nodal-chain phonons in K$_2$O.

\bibliographystyle{apsrev4-1}
\bibliography{ref}

\end{document}